# A Survey of RFID Authentication Protocols Based on Hash-Chain Method

Irfan Syamsuddin[a], Tharam Dillon[b], Elizabeth Chang[c], and Song Han[d]

[a]*State Polytechnic of Ujung Pandang, Indonesia*
[b,c,d]*DEBI Institute, Curtin University of Technology, Australia*
*irfans@ poliupg.ac.id, tharam@it.uts.edu.au, change@cbs.curtin.edu.au,
song.han@cbs.curtin.edu.au*

**Abstract**

*Security and privacy are the inherent problems in RFID communications. There are several protocols have been proposed to overcome those problems. Hash chain is commonly employed by the protocols to improve security and privacy for RFID authentication. Although the protocols able to provide specific solution for RFID security and privacy problems, they fail to provide integrated solution. This article is a survey to closely observe those protocols in terms of its focus and limitations.*

## 1. Introduction

Radio frequency identification or RFID was first used during the Second World War in Identification Friend or Foe systems onboard military aircraft. Soon after, Harry Stockman demonstrated a system energized completely by reflected power. Then in 1960s, the first Electronic Article Surveillance anti-theft systems were commercialized. Later, in 1970s, the US Department of Energy investigated the technology's potential to safeguard materials at nuclear weapons sites [17].

Recently, radio frequency identification (RFID) has been regarded as the main driver of the future ubiquitous technology. It is also claimed as the core technology to realize internet of things environment where large amount of items are connected seamlessly anytime and anywhere [2]. RFID offers simplicity for people to object (P2O) and object to object (O2O) communications. It is believed that it will play a significant role for future ubiquitous society [11].

Generally, RFID systems consist of Radio Frequency Identification (RFID) tags and RFID readers. While RF tags operate as transponders, RF readers act as transceivers. In case of a more complex application, a database server is required to store information comes from both transponders and receivers sides [12].

It is assumed that communication between RF tags and server is secure. The process of RFID communication can be described as follows, RFID reader request access to the RFID tag and return the reply to the database server. After identification and authentication on server side, then server will return the information of RFID tag to the reader [12][13].

Automatic identification is the basic characteristic of RFID. In its simplest form, identification can be binary, e.g., paid or nor paid which is useful for alerting. Therefore, alerting is become the next powerful feature of RFID. Also, RFID enable real time monitoring to a large number of in a short time. In addition, RFID has ability to perform on-chip computation, accordingly it support cryptographic protocol for authentication. In general, RFID has four basic capabilities, identification, alerting, monitoring, and authentication [9].

RFID has become a new and exciting area of technological development, and is receiving increasing amounts of attention. There is tremendous potential for applying it even more widely, and increasing numbers of companies have already started up pilot schemes or successfully used it in real-world environments.

Based on the various industry areas that are featured in the reviewed literature, RFID technology is widely used in many areas such as
 - Animal detection.
 - Aviation.
 - Transportation.
 - Building management.
 - Construction.

- Enterprise feedback control.
- Fabric and clothing.
- Food safety warranties.
- Health,
- Military, etc.

Therefore, RFID related business experiences many significant advantages. Das [21] confirms market research report from IDTechEx that the increasing sales of RFID tags for the 60 years up to the beginning of 2006 reached 2.4 billion, which was accounted for more than 600 million tags being sold in 2005. Then in 2006, it was expected that approximately 1.3 billion tags and 500 million RFID smart labels would be needed in a range of areas, such as retailing, logistics, animals and farming, library services, and military equipment.

In terms of academic point of view, RFID is regarded as an exciting research area due to its relative uniqueness and exploding growth. RFID research has led to the emergence of a new academic research area that builds on existing research in a host of disciplines, such as electronic engineering, information systems, computer science, environmental science, medical and public health and also business strategic management, and there has been a significant increase in the number of papers on RFID in research journals [20]. As noted by Heinrich [22], RFID is likely to be among the most exciting and fastest-growing technologies in terms of scope of application in the next generation of business intelligence which attract many researchers from different field to collaborate in a specific research.

However, to continue the advancement of knowledge in this area, it is important to understand the current status of RFID research and to examine contemporary trends in the research domain. It is vital to determine the principal concerns of current RFID research, whether technological, application related, or security related. An academic review of the literature is essential for appropriately shaping future research.

One of a growing area in RFID literature is authentication protocol with hash chain model which will be deeply discussed in this paper. The rest of this paper is organized as follows. In the next section, we will briefly review security and privacy issues of RFID. The application of hash chain in RFID authentication is presented in Sections 3. In Section 4, we propose our comparative study of recent RFID authentication protocols based on hash chain model. And some conclusions will be made in the last section.

## 2. Security and privacy of RFID

RFID technology poses exclusive privacy and security concerns since it is promiscuous the tags themselves typically maintain no history of past readings. As a result, security and privacy issues are considered as the fundamental issue the RFID technology [4][12].

### 2.1. Security Objectives

Although generally it is assumed that communication between RFID tag and the readers is secure, yet since it is basically wireless based communication, a number of security and privacy issues could not be avoided. Fundamental information security objectives, such as confidentiality, integrity, availability, authentication, authorization, non-repudiation and anonymity are often not achieved unless special security mechanisms are integrated into the system [23].

The privacy aspect has gained special attention for RFID systems. Consumers may carry objects with silently communicating transponders without even realising the existence of the tags. Passive tags usually send their identifier without further security verification when they are powered by electromagnetic waves from a reader. The ID information can also be linked to other identity data and to location information. Consumers might employ a personal reader to identify tags in their environment but the large number of different standards may render this difficult. Therefore, user privacy is the main consideration of RFID security.

### 2.2. Security Properties

A. Confidentiality
The communication between reader and tag is unprotected in most cases. Eavesdroppers may thus listen in if they are in immediate vicinity. Furthermore, the tag's memory can be read if access control is not implemented.

B. Integrity
With the exception of high-end ISO 14443 systems which use message authentication codes (MACs), the integrity of transmitted information cannot be assured. Checksums (CRCs) are often employed on the communication interface but protect only against random failures. Furthermore, the writable tag memory can be manipulated if access control is not implemented.

C. Availability

Any RFID system can easily be disturbed by frequency jamming. But, denial-of-service attacks are also feasible on higher communication layers.

D. Authenticity

The authenticity of a tag is at risk since the unique identifier (UID) of a tag can be spoofed or manipulated. The tags are in general not tamper resistant.

## 2.3. RFID Security and Privacy Schema

To get a clear understanding on how those security and privacy issues exist within RFID technology, a schema introduced in [17] is presented.

Garkenfield, *et.al*. [17] clearly describe the actual existing security and privacy problems within RFID technology as can be seen in figure 1.

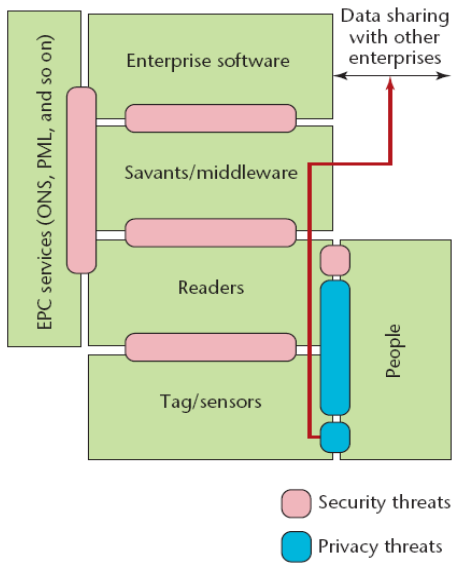

**Figure 1.** Security and Privacy threats within RFID

They use the Electronic Product Code (EPC) based schema as a popular account of RFID technology. It shows where security and privacy threats exist within common RFID applications.

Physical attacks, denial of service, man in the middle attack, eavesdropping, traffic analysis, counterfeiting, and tag cloning attack are several security related problems commonly addressed to RFID technology [10][17].

In terms of privacy, RFID is mainly questioned in tracking and inventorying capabilities. Another privacy concern is authentication problem. Authenticating legal communication between RFID tag and reader is the main question intensive studies in this field [10][12][13].

This paper surveys some papers that introduce RFID authentications with hash chain method.

## 3. Hash Chain Model and Its applications in RFID authentication

Hash chain is basically a cryptography approach for safeguarding against password eavesdropping which is firstly proposed in [7]. Now, it can be found in other applications such as micropayment systems and RFID authentication due to elegant and versatile low-cost associated to this technique.

Lamport [7] describes that a hash chain of length N could be constructed by applying a one-way hash function h(.) recursively to an initial seed value s.

$$h^N(s) = h(h(...h(s)...)) \text{ (N times)}$$

The last element hN(s) is also called the tip T of the hash chain. By knowing $h^N(s)$, $h^{N-1}(s)$ can not be generated by those who do not know the value *s*, however given $h^{N-1}(s)$, its correctness can be verified using $h^N(s)$. This property of hash chains has evolved from the property of one-way hash functions [7].

Then, the application of hash chain model into RFID application is firstly introduced by Ohkubo et.al [8]. After reviewing previous protocols for improving privacy of RFID applications, they suggest five points for an approach to RFID scheme design.
 - keep complete user privacy.
 - eliminate the need for extraneous rewrites of the tag information.
 - minimize the tag cost.
 - eliminate the need for high power of computing units.
 - provide forward security.

Their protocol for secure RFID privacy protection scheme is described as follows [8].

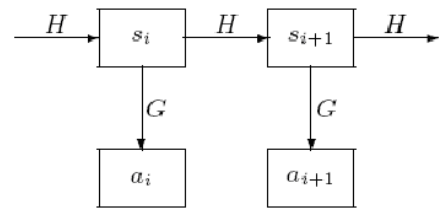

**Figure 2.** Hash chain in RFID application.

Hash chain technique is employed to renew the secret information contained in the tag from G to H. The following is a brief description how it works. In

561

Authorized licensed use limited to: Seoul National University. Downloaded on August 14,2010 at 15:00:52 UTC from IEEE Xplore. Restrictions apply.

the beginning, a tag has initial information $s_1$. In the $i$-th transaction with the reader, the RFID tag will do two things:

1. Sends answer $a_i = G(s_i)$ to the reader,
2. Renews secret $s_{i+1} = H(s_i)$ as determined from previous secret $s_i$,

where $H$ and $G$ are hash functions, as in Figure 1. The reader sends $a_i$ to the back-end database.

The back-end database maintains a list of pairs (*ID*, $s_i$), where $s_i$ is the initial secret information and is different for each tag. So the back-end database that received tag output $a_i$ from the reader calculates $a0i = G(Hi(s_i))$ for each $s_i$ in the list, and checks if $a_i \stackrel{?}{=} a'_i$. Then, it will find $a'_i$, $a'_i = a_i$, then return the ID, which is a pair of $a'_i$.

This is the basic hash chain method implementing in RFID authentication protocol which is then followed by many researchers by proposing new protocols in different perspectives and techniques.

## 4. Comparative study of hash chain based RFID authentication protocols

This part will discuss several RFID authentication protocols that using hash chain as a method for properly enhancing security and privacy of RFID. We introduce hash chain method of each protocol and review its limitation. There are ten protocols of RFID authentication with hash chain will be compared in this section.

As mentioned earlier, Ohkubo et al. [8] proposed a hash-based authentication protocol. The aim of the protocol is to provide better protection of user privacy with the basic concept of refreshing the identifier of the tag each time it is queried by a reader. The protocol changes RFID identities on each read based on hash chains. Hash chain method is used in this two ways communication of RFID tag. This protocol does not require a random number generator. However, it is confirmed that this protocol is flawed to certain replay attacks.

The next protocol is a cryptographically controlled access protocol for RFID tag by using hash locks was proposed by Weis et al. [15]. They argue that although the hash value can be read out by any reader, yet only authorized ones would be able to look up the tags key in a database of key-hash pairs. The objective of this protocol is improving RFID tag security and privacy by using an integrated hash function where key could be verified by comparing the key hash with the stored hash value. The drawback of this protocol is that the static hash value would still be traceable.

In addition, Molnar et al. [4] proposed a hash-tree based authentication protocol for RFID tags. They exposed privacy issues related to RFID in libraries, described current deployments, and suggested novel architectures for library RFID. This protocol utilizes a dynamic amount of computation required per tag, which depends on the number of tags available in the hash-tree. However, the protocol is confirmed has a serious problem, where another security leak will arise if the tag lost. In this case, anonymity for the rest of the hash-tree group may be compromised by attacker. Therefore, the protocol does not provide for forward-anonymity.

The following approach is the hash chain based RFID authentication protocol by Henrici [3]. This protocol only requires a hash function in the tag and data management at the back-end. It offers a high degree of location privacy and is resistant to many forms of attacks. Further, only a single message exchange is required, the communications channel needs not be reliable and the reader/third party need not be trusted, and no long-term secrets need to be stored in tags. However, the limitation of this protocol is that it does not guarantee to provide full privacy, since the tag is vulnerable to tracing when the attacker interrupts the authentication protocol mid-way. Therefore, this approach also has

Then, Avoine [5] introduced the modification of authentication protocol. This proposed protocol is aimed at solving the replay attack problem of [8]. He argued that privacy issues cannot be solved without looking at each layer separately, where RFID system has three layers, application, communication and physical. Yet, it does not consider the issue of availability, and the protocol is vulnerable to attacks where the attacker forces an honest tag to fall out of synchronization with the server so that it can no longer authenticate itself successfully.

Similarly, anonymous RFID protocol is also offered by Dimitriou [19]. The objective of this protocol is to protecting forward privacy from cloning and privacy attacks. Mutual authentication is performed within the protocol based on the use of secrets which are shared between tag and database, and refreshed to avoid tag tracing. Yet, limitation of the protocol is addressed to desynchronization problem which occurs in database side. This is confirmed to be vulnerable from man in the middle attack.

The following protocol is offered by Rhee et al. [6]. It is called hash-based challenge-response which is aimed at providing security protection mechanism from the replay and spoofing attacks. The proposed protocol is based on challenge response using one-way hash function and random number which is claimed suitable for security database environment. Hash chain function is used in the protocol to guarantee secret key in the form of ID. Then, the tag does not need to update the secret key which avoids attacks by interrupting the session. However, this solution does not provide forward secrecy which means if a tag can be compromised then attacker will be able to trace the past communications from the same tag.

Lee et al. [18] proposed a new RFID authentication protocol with hash chain. The objective of this effort is to solve the desynchronization problem by maintaining a previous identification number in the database server. However, since the hashed identification number is always identical, an adversary who queries tag actively without updating identity able to trace the RFID tag. Unfortunately, although it able to secure desynchronization problem, this protocol still suffer from traceability attack which is become a serious limitation of the protocol to solve privacy problem of RFID.

Likewise, RFID authentication scheme whith a hash function and synchronized secret information was introduced by Lee, *et.al.* [14]. The protocol is aimed at securing user privacy including against tag cloning attack through an additional hash operation. Unfortunately, this protocol suffers from desynchronization attacks that could be conducted by adversaries. This occurs due to unavailability of PRNG in the RFID tag while the server does not know how many times an RFID tag may have not yet updated its secret information.

Finally, Han, *et.al.* [13] offer new kind of mutual authentication protocol to solve some problems of previous protocols. In their protocol, the authentication mechanism is supported by a monitoring component. The component which exists in database server, constantly monitors the synchronized secret information between RFID tag and reader. This protocol is argued to provide more secure communication mechanism since the communication between tag, reader and database is mutually authenticated and constantly monitored. In addition, the protocol also supports the low-cost non-volatile memory of RFID tags. However, it also has limitation since it still needs the back-end database support.

**Table 1.** Comparison of all RFID authentication protocols with hash chain method.

| AUTHOR | Hash Chain | Privacy | Anonymity |
|--------|-----------|---------|-----------|
| Ohkubo | Yes | Yes | Yes |
| Weis | Yes+PRNG | Yes | Yes |
| Molnar | Yes | No | Yes |
| Henrici | Yes | Yes | Yes |
| Avoine | Yes | Yes | Yes |
| Dimitriou | Yes+PRNG | Yes | Yes |
| Rhee | Yes+PRNG | Yes | Yes |
| Lee | Yes | Yes | Yes |
| Lee | Yes+PRNG | Yes | Yes |
| Han | Yes | Yes | Yes |

As can be seen in table 1, all protocols being studied are vary in terms of the way of using hash chain method but focus on the same objective that is providing better privacy mechanism while maintaining anonymity.

## 5. Conclusion

RFID authentication protocols in this study provide privacy and anonymity. Hash chain method is used in these RFID authentication protocols in various ways a unique solution for security and privacy problems of RFID technology. As a result, while problems in particular cases can be addressed, other problem is arising. Therefore, it can be concluded that recent RFID authentication protocols with hash chain failed to satisfy an integrated security and privacy solutions for RFID.

Based on the findings, we recommend future study in this area to focus on a level such as developing a framework for RFID authentication model. The framework will be useful for researchers as a foundation for collaborative research in the future for integrated RFID authentication solutions.